\pdfoutput=1 

\documentclass[iop]{emulateapj}
\usepackage{epsfig}
\usepackage{apjfonts}
\usepackage{aas_macros}

\begin{document}

\title{Limits on the Weak Equivalence Principle and Photon Mass with FRB 121102 Sub-pulses}
\author{Nan Xing$^{1}$, He Gao$^{1,*}$, Junjie Wei$^{2}$, Zhengxiang Li$^{1}$, Weiyang Wang$^{3,4}$, Bing Zhang$^{5}$, Xue-Feng Wu$^{2}$ and Peter M{\'e}sz{\'a}ros$^{6,7,8}$}
\affiliation{
$^1$Department of Astronomy, Beijing Normal University, Beijing 100875, China; gaohe@bnu.edu.cn\\
$^2$Purple Mountain Observatory, Chinese Academy of Sciences, Nanjing 210008, China;\\
$^3$Key Laboratory for Computational Astrophysics, National Astronomical Observatories, Chinese Academy of Sciences, 20A Datun Road, Beijing 100101, China;\\
$^4$University of Chinese Academy of Sciences, Beijing 100049, China;\\
$^5$Department of Physics and Astronomy, University of Nevada Las Vegas, NV 89154, USA\\
$^6$ Department of Astronomy and Astrophysics, Pennsylvania State University, 525 Davey Laboratory, University Park, PA 16802, USA\\
$^7$ Department of Physics, Pennsylvania State University, 104 Davey Laboratory, University Park, PA 16802, USA\\
$^8$ Center for Particle and Gravitational Astrophysics, Institute for Gravitation and the Cosmos, Pennsylvania State University, 525 Davey Laboratory, University Park, PA 16802, USA. 
  }

\begin{abstract}
Fast radio bursts (FRBs) are short duration ($\sim$millisecond) radio transients with cosmological origin. The simple sharp features of the FRB signal have been utilized to probe two fundamental laws of physics, namey, testing Einstein's weak equivalence principle and constraining the rest mass of the photon. Recently, \cite{hessels18} found that after correcting for dispersive delay, some of the bursts in FRB 121102 have complex time-frequency structures that include sub-pulses with a time-frequency downward drifting property. Using the delay time between sub-pulses in FRB 121102, here we show that the parameterized post-Newtonian parameter $\gamma$ is the same for photons with different energies to the level of $\left|\gamma_1-\gamma_2\right|<2.5\times 10^{-16}$, which is 1000 times better than previous constraints from FRBs using similar methods. We also obtain a stringent constraint on the photon mass, $m_{\gamma} < 5.1\times10^{-48}$ g, which is 10 times smaller than previous best limits on the photon mass derived through the velocity dispersion method. 
\end{abstract}



\section{INTRODUCTION}

Fast radio bursts (FRBs) are short duration radio transients with anomalously high dispersion measure in excess of the Galactic value (DM$\gtrsim200\,{\rm pc\,cm^{-3}}$; \cite{lorimer07,keane12,thornton13,petroff16}). The first repeating burst FRB 121102, was localized in a star-forming dwarf galaxy at $z = 0.193$, which has confirmed the cosmological origin of FRBs \citep{spitler16,scholz16,chatterjee17,marcote17,tendulkar17}. Although the progenitors and radiation mechanism are still debated, FRBs have been proposed to be promising tools for cosmological and astrophysical studies, such as locating the ``missing'' baryons \citep{mcquinn14},  constraining the cosmological parameters \citep{gao14, zhou14, yang16a, walters18}, directly measure $\Omega_b$ of the universe \citep{deng14,keane16} and probe the reionization history of the universe \citep{deng14,zheng14,caleb19,li19}, probing compact dark matter or precisely measuring the Hubble constant and the cosmic curvature through gravitationally lensed FRBs \citep{munoz16,wang18,li18}, measuring cosmic proper distances \citep{yu17}, testing the Einstein's weak equivalence principle \cite[WEP,][]{wei15, nusser16,tingay16,wu17,yu18} and constraining the rest mass of the photon \citep{wu16,bonetti16,bonetti17,shao17}.

FRB emission arrives later at lower radio frequencies. In principle, the observed time delay for a cosmic transient between two different energy bands should include various terms \citep{gao15,wei15}, such as the intrinsic (astrophysical) time delay $\Delta t_{\rm int}$, the time delay contribution from the dispersion by the line-of-sight free electron content $\Delta t_{\rm DM}$, the potential time delay caused by special-relativistic effects ($\Delta t_{\rm spe}$) in the case where the photons have a rest mass which is non-zero, and the potential time delay caused by the violation of Einstein's weak equivalence principle ($\Delta t_{\rm gra}$) where photons with different energies following different trajectory while traveling in the same gravitational potential. In FRB observations, the arrival time delay $\Delta t_{\rm obs}$ is around 1s and at a given frequency $\nu$ follows a $\nu^{-2}$ law \citep{lorimer07,keane12,thornton13,petroff16}, indicating that the time delay should mainly be attributed to dispersive delay $\Delta t_{\rm DM}$. Even if the WEP is indeed violated or if the rest mass of the photon is indeed nonzero, the contribution of $\Delta t_{\rm gra}$ and $\Delta t_{\rm spe}$ to $\Delta t_{\rm obs}$ should be small. 

In this case, a conservative constraint on the WEP can be obtained under the assumption that $\Delta t_{\rm obs}$ is mainly contributed by $\Delta t_{\rm gra}$. Using FRB 110220 and two possible FRB/gamma-ray burst (GRB) association systems (FRB/GRB 101011A and FRB/GRB 100704A), \cite{wei15} obtained a strict upper limit on the differences of the parametrized post-Newtonian (PPN) parameter $\gamma$ values, e.g. $\left|\gamma(1.23\; \rm GHz)-\gamma(1.45\; \rm GHz)\right|<4.36\times10^{-9}$. \cite{keane16} reported 
the connection between a fading radio transient with FRB 150418, so that a putative host galaxy with redshift $0.492\pm0.008$ was identified (see counter opinions in \cite{williams16}, where the counterpart radio transient is claimed to be AGN variability instead of an afterglow of FRB 150418). 
Assuming that 0.492 is the redshift of FRB 150418, \cite{tingay16} and \cite{nusser16} obtained more stringent upper limits on the differences of $\gamma$ values as (1--2)$\times10^{-9}$ and $10^{-12}$--$10^{-13}$, by considering the Milky Way and the Large-scale structure gravitational potential respectively. 

On the other hand, if one assumes that $\Delta t_{\rm obs}$ of an FRB is mainly contributed by $\Delta t_{\rm spe}$, a conservative limit on the rest mass of the photon could be placed. For instance, taking $z=0.492$ as the redshift of FRB 150418, a conservative upper limit for the rest mass of the photon was placed as $m_{\rm ph} \leq 5.2 \times 10^{-47}$ g, which is $10^{20}$ times smaller than the rest mass of electron \citep{wu16,bonetti16}. Later, \cite{bonetti17} applied the similar method to FRB 121102, and they obtained a similar result as $m_{\rm ph} \leq 3.6 \times 10^{-47}$ g. 

Most recently, \cite{hessels18} found that some bursts in FRB 121102 have complex time-frequency structures that include subbursts with finite bandwidths. After correcting for dispersive delay, the subbursts still show an interesting sub-pulse time-frequency downward drifting pattern, namely the characteristic frequencies for sub-pulses drift lower at later times in the total burst envelope. The same features are also found in the second discovered repeating FRB source, FRB 180814.J0422+73 \citep{chime19}. Such features could be intrinsic [e.g. related to the burst emission mechanism \citep{wang19}], or they could also be imparted by propagation effects \citep{cordes17,hessels18}. Plasma lensing may cause upward and downward sub-pulse drifts, but only downward drifting is observed so far in the repeating FRBs. In the 1.1-1.7 GHz band, the typical time spans for sub-pulses are $\sim0.5-1$ ms, with a characteristic drift rate of $\sim200~\rm MHz~ms^{-1}$ toward lower frequencies. Considering that FRB 121102 is the only FRB with confirmed redshift observations, and the time delay between sub-pulses is almost $10^4$ times smaller than the dispersive delay, it is of great interest to place limits on the WEP and the photon mass with the FRB 121102 sub-pulses. 

\begin{table*}
{\footnotesize
\caption{Upper limits on the differences of the $\gamma$ values through the Shapiro time delay effect.}
\begin{tabular}{l|lllcc}
\hline\hline
&&&&& \\
{Categorization}&{Author (year)}&{Source}&{Messengers}&{Gravitational Field}&{$\Delta\gamma$}\\
&&&&& \\
\hline
           & This work                  & FRB 121102 & 1.374--1.344 GHz photons & Laniakea supercluster of galaxies & $2.5\times10^{-16}$ \\
            & \cite{wei15} & FRB 110220 & 1.2--1.5 GHz photons & Milky Way & $2.5\times10^{-8}$ \\
    &                            & FRB/GRB 100704A & 1.23--1.45 GHz photons & Milky Way & $4.4\times10^{-9}$ \\
 & \cite{tingay16} & FRB 150418 & 1.2--1.5 GHz photons & Milky Way & (1--2)$\times10^{-9}$ \\
    & \cite{nusser16} & FRB 150418 & 1.2--1.5 GHz photons & Large-scale structure & $10^{-12}$--$10^{-13}$ \\
          &     \cite{longo88}               & SN 1987A & 7.5--40 MeV neutrinos & Milky Way & $1.6\times10^{-6}$  \\
   ~Same particles                          & \cite{gao15}  & GRB 090510 & MeV--GeV photons & Milky Way & $2.0\times10^{-8}$ \\
~~~~~~~~with     &                  & GRB 080319B & eV--MeV photons & Milky Way & $1.2\times10^{-7}$ \\
     ~different energies                      & \cite{yang16b}  & Crab pulsar & 8.15--10.35 GHz photons  & Milky Way  & (0.6--1.8)$\times10^{-15}$ \\
& \cite{zhanggong16}  & Crab pulsar & eV--MeV photons  & Milky Way  & 3.0$\times10^{-10}$ \\
       & \cite{leung18}  & Crab pulsar & 1.52--2.12 eV photons  & Milky Way  & 1.1$\times10^{-10}$ \\
  & \cite{wei16b}  & Mrk 421 & keV--TeV photons & Milky Way & $3.9\times10^{-3}$ \\
 &                             & PKS 2155-304 & sub TeV--TeV photons & Milky Way & $2.2\times10^{-6}$ \\
                 & \cite{wu16b}  & GW150914 & 35--150 Hz GW signals  & Milky Way  & $\sim10^{-9}$ \\
       & \cite{kahya16}  & GW150914 & 35--250 Hz GW signals  & Milky Way  & 2.6$\times10^{-9}$ \\
\hline
          & \cite{krauss88} & SN 1987A & eV photons and MeV neutrinos & Milky Way & $5.0\times10^{-3}$ \\
& \cite{longo88} & SN 1987A & eV photons and MeV neutrinos & Milky Way & $3.4\times10^{-3}$ \\
& \cite{wei16a}  & GRB 110521B & keV photons and TeV neutrino  & Laniakea supercluster of galaxies & $1.3\times10^{-13}$ \\
& \cite{wang16}  & PKS B1424-418 & MeV photons and PeV neutrino & Virgo Cluster & $3.4\times10^{-4}$ \\
       &                             & PKS B1424-418 & MeV photons and PeV neutrino & Great Attractor & $7.0\times10^{-6}$ \\
  Different particles            & \cite{boran19}  & TXS 0506+056 & GeV photons and TeV neutrino & Milky Way & $5.5\times10^{-2}$ \\
   & \cite{wei19b}  & TXS 0506+056 & GeV photons and TeV neutrino & Laniakea supercluster of galaxies & $10^{-6}$--$10^{-7}$ \\
    & \cite{wei17}  & GW170817 & MeV photons and GW signals  & Virgo Cluster  & 9.2$\times10^{-11}$ \\
       &                             & GW170817 & eV photons and GW signals   & Virgo Cluster  & 2.1$\times10^{-6}$ \\
       & \cite{abbott17}  & GW170817 & MeV photons and GW signals  & Milky Way   & -2.6$\times10^{-7}$---1.2$\times10^{-6}$ \\
       & \cite{shoemaker18}  & GW170817 & MeV photons and GW signals  & Milky Way  & 7.4$\times10^{-8}$ \\
\hline
       & \cite{wu17}  & GRB 120308A & Polarized optical photons  & Laniakea supercluster of galaxies  & $1.2\times10^{-10}$ \\
~Same particles         &                             & GRB 100826A & Polarized gamma-ray photons  & Laniakea supercluster of galaxies  & $1.2\times10^{-10}$ \\
~~with different   & & FRB 150807 & Polarized radio photons  & Laniakea supercluster of galaxies  & $2.2\times10^{-16}$ \\
 polarization states    & \cite{yang17}  & GRB 110721A & Polarized gamma-ray photons  & Milky Way  & $1.6\times10^{-27}$ \\
    & \cite{wei19}  & GRB 061122  & Polarized gamma-ray photons  & Laniakea supercluster of galaxies  & $0.8\times10^{-33}$ \\
       &                             & GRB 110721A & Polarized gamma-ray photons  & Laniakea supercluster of galaxies  & $1.3\times10^{-33}$ \\

\hline\hline
\end{tabular}
\label{table1}
}
\end{table*}

\section{Testing the Einstein weak equivalence principle}

The Einstein weak equivalence principle is an important foundation of many metric theories of gravity, including general relativity. One statement of the WEP is that test particles traveling in the same gravitational potential will follow the same trajectory, regardless of their internal structure and composition \citep{will06}. Therefore, it has long been proposed that the accuracy of the WEP can be constrained with the time delay for different types of messenger particles (e.g. photons, neutrinos, or gravitational waves), or the same types of particles but with different energies or different polarization states, which are simultaneously radiated from the same astronomical sources.

According to the Shapiro time delay effect \citep{shapiro64}, the time interval required for test particles to traverse a given distance would be longer by
\begin{equation}
t_{\rm gra}=-\frac{1+\gamma}{c^3}\int_{r_e}^{r_o}~U(r)dr\;,
\end{equation}
in the presence of a gravitational potential $U(r)$, where the test particles are emitted at $r_{e}$ and received at $r_{o}$. Here $\gamma$ is one of the parametrized post-Newtonian (PPN) parameters, which reflects how much space curvature is produced by unit rest mass. When the WEP is invalid, different particles might correspond to different $\gamma$ value. In this case, two particles emitted simultaneously from the source will arrive at the Earth with a time delay difference
\begin{equation}
\Delta t_{\rm gra}=\frac{\gamma_{\rm 1}-\gamma_{\rm 2}}{c^3}\int_{r_o}^{r_e}~U(r)dr\;,
\label{gra}
\end{equation}
where $\gamma_{\rm 1}$ and $\gamma_{\rm 2}$ correspond to two different test particles. For a cosmic source, in principle, $U(r)$ has contributions from the host galaxy potential $U_{\rm host}(r)$, the intergalactic potential $U_{\rm IG}(r)$ and the local gravitational potential $U_{\rm local}(r)$. Since the potential models for $U_{\rm IG}(r)$ and $U_{\rm host}(r)$ are poorly known, for the purposes of obtaining a lower limit, it is reasonable to extend the local potential out to cosmic scales to bracket from below the potential function of $U_{\rm IG}(r)$ and $U_{\rm host}(r)$. In the previous works, the gravitational potential of the Milky Way (MW) or the Laniakea supercluster \citep{tully14} has been used as the local potential, which could be expressed as a Keplerian potential \footnote{Although the potential models for the Laniakea supercluster is still not well known, it has been tested that the adoption of the Keplerian potential model, comparing with other widely used potential model, i.e., the isothermal potential would not have a strong influence on the results for testing the WEP \citep{krauss88}.} $U(r)=-GM/r$. In this case, we have 
\begin{eqnarray}
\Delta t_{\rm gra}= \left(\gamma_{1}-\gamma_{2}\right ) \frac{GM}{c^{3}} \times \qquad\qquad\qquad\qquad\qquad\\ \nonumber
\ln \left\{ \frac{ \left[d+\left(d^{2}-b^{2}\right)^{1/2}\right] \left[r_{L}+s_{\rm n}\left(r_{L}^{2}-b^{2}\right)^{1/2}\right] }{b^{2}} \right\}\;,
\label{eq:gammadiff}
\end{eqnarray}
where $d$ is the distance from the transient to the MW/Laniakea center and $b$ is the impact parameter of the light rays relative to the center. Here we use $s_{\rm n}=+1$ or $s_{\rm n}=-1$ to denote the cases where the transient is located along the direction of the MW/Laniakea or anti-MW/Laniakea center. For a cosmic source, $d$ is approximated as the distance from the source to the Earth. The impact parameter can be estimated as 
\begin{equation}
b=r_{L}\sqrt{1-(\sin \delta_{s} \sin \delta_{L}+\cos \delta_{s} \cos \delta_{L} \cos(\beta_{s}-\beta_{L}))^{2}}\;,
\end{equation}
where $\beta_{s}$ and $\delta_{s}$ are the source coordinates, $\beta_{L}$ and $\delta_{L}$ represent the coordinates of the local (MW/Laniakea) center, and $r_{L}$ is the distance from the Earth to the center. 

In the literature, many investigations have been done to achieve stringent limits on $\gamma$ differences between particles emitted from the same astrophysical sources, such as supernovae 1987A \citep{krauss88,longo88}, GRBs \citep{gao15,wei16a,wu17,yang17,wei19}, FRBs \citep{wei15,tingay16,nusser16,wu17}, blazars \citep{wei16b,wang16,wei19b}, the Crab pulsar \citep{yang16b,zhanggong16}, and gravitational wave (GW) sources \citep{wu16b,kahya16,abbott17,shoemaker18,wei17}. The previous results are summarized in Table 1. When the test particles are of different species, the best constraint is $\left|\gamma_1-\gamma_2\right|<1.3\times 10^{-13}$ for keV photons and TeV neutrino from GRB 110521B \citep{wei16a}. When the test particles are the same species but with different energies, the best constraint is $\left|\gamma_1-\gamma_2\right|<(0.6-1.8)\times 10^{-15}$ for 8.15-10.35 GHz photons from the Crab pulsar \citep{yang16b}. When the test particles are of the same species but with different polarization states, the best constraint is $\left|\gamma_1-\gamma_2\right|<0.8\times 10^{-33}$ for polarized gamma-ray photons from GRB 061122 \citep{wei19}.

Here we considered the time-frequency structure of FRB 121102. As shown in \cite{hessels18}, some of FRB 121102 repeated bursts have several sub-pulses, which have higher frequencies arriving earlier. We consider the closest neighboring sub-pulses in AO-05, where the time delay between $f_1=1374.16$ MHz and $f_2=1343.69$ MHz is 0.4 ms. With the inferred coordinates and redshifts for FRB 121102 [here we adopt $\rm R.A.=\beta_{s}=05^h32^m$, $\rm Dec.=\delta_{s}=+33^{\circ}08'$ and $z = 0.193$ \citep{spitler16}], a stringent limit on the WEP can be placed as 
\begin{equation}
\left|\gamma_1-\gamma_2\right|<2.5\times 10^{-16},
\end{equation}
where we consider the gravitational potential of the Laniakea supercluster as the local potential, Great Attractor ($\beta_{L}=10^{\rm h}32^{\rm m}$, $\delta_{L}=-46^{\circ}00^{'}$) is adopted as the gravitational center of Laniakea \citep{lynden-Bell88}, $M_{\rm L}\simeq10^{17}M_{\odot}$ is the Laniakea mass and $r_{L}=79$ Mpc is the distance from the Earth to the center of Laniakea \citep{tully14}. The result is 1000 times better than previous constraints from FRBs and 4 times better than previous best constraints for the case when the test particles are of the same species but with different energies.

\section{Constraints on the photon mass} \label{sec2}

The postulate that all electromagnetic radiation propagates in vacuum at the constant speed $c$, namely that the photons should have a zero rest mass, is one of the most important foundations of Einstein's theory of special relativity. If the photon mass is nonzero, a mass term should be added to the Lagrangian density for the electromagnetic field to describe the effective range of the electromagnetic interaction \citep{proca36}. In this case, some abnormal phenomena will appear for the electromagnetic potentials and their derivatives, for instance, the speed of light is no longer constant but depends on the frequency of the photons, magnetic dipole fields would decrease with distance very rapidly due to the addition of a Yukawa component, longitudinal electromagnetic waves could exist, and so on. Such effects could be applied to make restrictive constraints on the photon rest mass \citep{goldhaber71, tu05, pani12}. For instance, it has long been proposed that the photon rest mass could be constrained by using the frequency-dependent time delays of multi-wavelength emissions from astrophysical sources \citep{lovell64,warner69,schaefer99,wu16,bonetti16,bonetti17,shao17,wei18}.  

According to Einstein's special relativity, if the photon has a rest mass $m_{\gamma}$, the photon energy can  be written as 
\begin{equation}\label{eq1}
E=h\nu=\sqrt{p^2c^2+m_\gamma^2c^4}\;,
\end{equation}
where $h$ is the Planck constant. In vacuum, the speed of photons with energy $E$ can be derived as
\begin{equation}
\upsilon=\frac{\partial{E}}{\partial{p}}\;.
\end{equation}
When $m_{\gamma}=0$, we have $\upsilon=c$. If $m_{\gamma}\neq0$, we have 
\begin{equation}\label{eq2}
\upsilon=\frac{\partial{E}}{\partial{p}}=c\sqrt{1-\frac{m_\gamma^2c^4}{E^2}}\approx c\left(1-\frac{1}{2}\frac{m_\gamma^2c^4}{h^2\nu^2}\right)\;,
\end{equation}
where the last approximation is applicable when $m_\gamma\ll h\nu/c^{2}\simeq7\times10^{-39}\left(\frac{\nu}{\rm GHz}\right)\;{\rm g}$. In this case, two photons with different frequencies, which are emitted simultaneously from the same source, would arrive on the Earth at different times with a time-frequency downward drifting pattern. For a cosmic source, the arrival time difference is given by
\begin{equation}\label{eq5}
  \Delta{t_{m_{\gamma}}}=\frac{m_\gamma^2c^4}{2h^2H_0}\left(\nu_l^{-2}-\nu_h^{-2}\right)\int_0^{z}\frac{(1+z')^{-2}dz'}{\sqrt{\Omega_{\rm m}(1+z')^3+\Omega_{\Lambda}}}\;,
\end{equation}
where $H_0$ is the Hubble constant. Thus, the photon mass can be constrained as \citep{wu16}
\begin{equation}\label{eqmgamma1}
m_{\gamma}=\left(1.54\times10^{-47}\rm g\right)
\left[\frac{\Delta{t_{m_{\gamma}}}}{\left(\nu_{l,9}^{-2}-\nu_{h,9}^{-2}\right)\int_0^{z}\frac{(1+z')^{-2}dz'}{\sqrt{\Omega_{\rm m}(1+z')^3+\Omega_{\Lambda}}}}\right]^{1/2}\;,
\end{equation}
where $\nu_{9}$ is the radio frequency in units of $10^{9}$ Hz. 

In the literature, many attempts have been made to obtain constraints on the photon rest mass by considering various astrophysical sources, including flare stars \citep{lovell64}, the Crab Nebula pulsar \citep{warner69}, FRBs \citep{wu16,bonetti16,bonetti17,shao17}, GRBs \citep{schaefer99} and pulsars in the Large and Small Magellanic Clouds \citep{wei18}. The constraint results are shown in Figure 1. The current best constraint on the photon mass through the velocity dispersion method is made by using the radio emissions from FRB 121102, $m_{\gamma} \leq 3.6 \times 10^{-47}$ g \citep{bonetti17}, where the time delay between the whole observational bandwidth is considered, and $\Delta t_{\rm obs}$ is in order of 1 second. 

Here we propose to use the observed time delay between sub-pulses in FRB 121102, such as the closest neighboring sub-pulses in AO-05 ($\Delta t_{\rm obs}=0.4$ ms between $f_1=1374.16$ MHz and $f_2=1343.69$ MHz) to obtain more stringent constraints on the photon mass as $m_{\gamma} < 5.1\times10^{-48}$ g, where $z = 0.193$ is adopted for FRB 121102, and the Planck results are adopted for cosmological parameters, e.g. $H_0=67.8{\rm km~s^{-1}Mpc^{-1}}$, $\Omega_m=0.308$ and $\Omega_\Lambda=0.692$ \citep{planck}. 

As shown in Figure 1, our result is 10 times better than that obtained using other FRB sources, and $\sim10^{4}$ times better than that obtained by GRBs, $\sim10^{3}$ times better than that obtained by pulsars in the Large and Small Magellanic Clouds, $\sim10^{6}$ times better than flare stars and $\sim10^{7}$ times better than the Crab Nebula pulsar. 

\begin{figure}[h]
\vskip-0.2in
\centerline{\includegraphics[angle=0,scale=0.6]{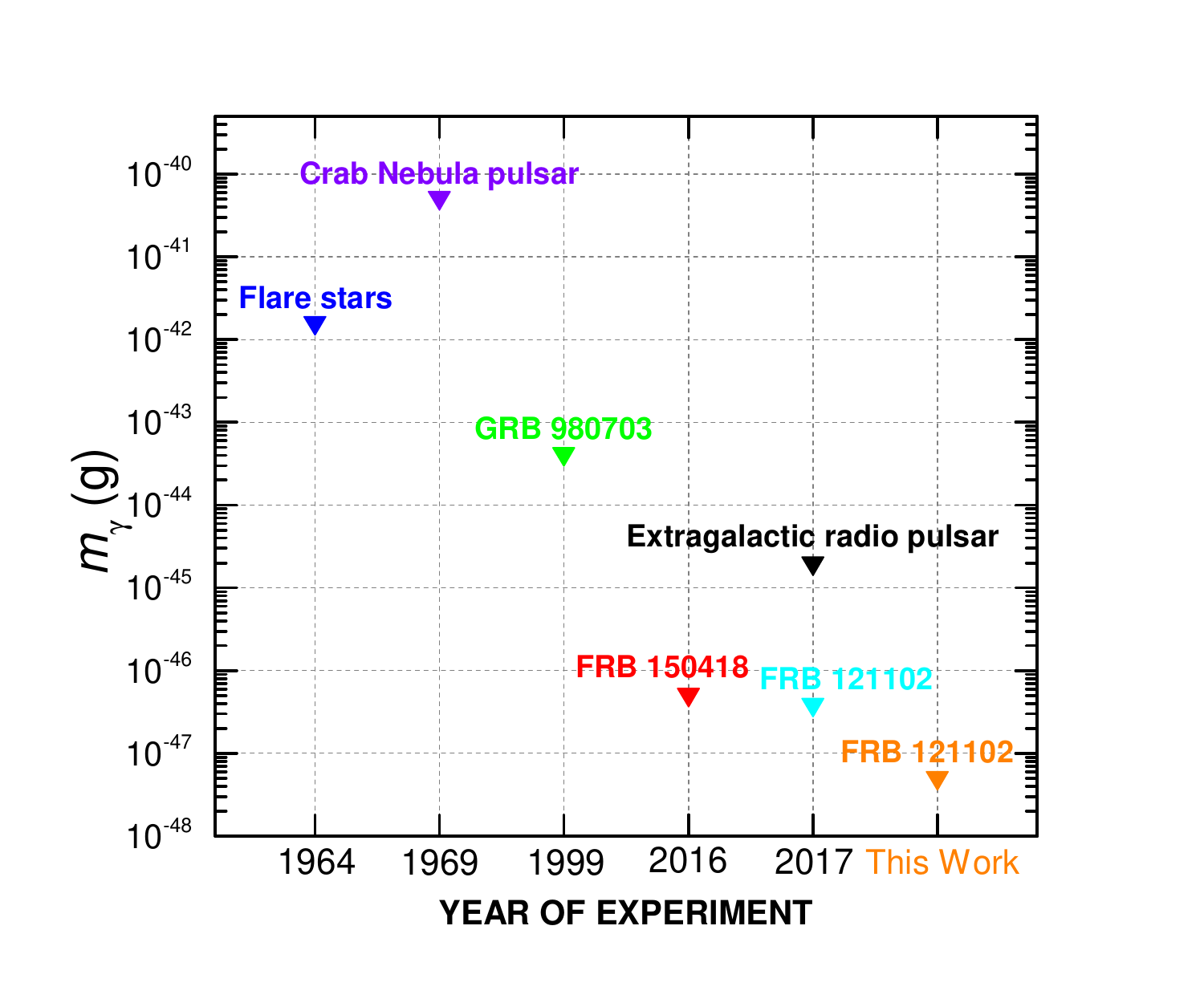}}
\vskip-0.1in
\caption{Strict upper limits on the photon rest mass from the velocity dispersion method, including the upper limits from flare stars \citep{lovell64}, Crab nebula pulsar \citep{warner69} and GRB 980703 \citep{schaefer99}, extragalacitc radio pulsars \citep{wei18}, FRB 150418 \citep{wu16,bonetti16}, FRB 121102 \citep{bonetti17} and FRB 121102 sub-pulses.}\label{f1}
\end{figure}

\section{Discussion}

Using the time-frequency structure of sub-pulses in some bursts of FRB 121102, here we have obtained a stringent limit on the $\gamma$ differences between photons with different energies of $\left|\gamma_1-\gamma_2\right|<2.5\times 10^{-16}$, which is 1000 times better than previous constraints from FRBs through similar methods. In addition, we also obtained a stringent constraint on the photon mass of $m_{\gamma} < 5.1\times10^{-48}$ g, which is 10 times better than the previous best limits on the photon mass using the velocity dispersion method. 

It is worth stressing the advantages of the method for placing limits on the WEP and the photon mass using the time-frequency structure of the sub-pulses of, e.g., FRB 121102. In previous works, the time delay between the whole observational bandwidth of FRBs (in order of 1 s) were applied to make constraints on the WEP or the photon mass. It is clear that such a time delay should mainly be attributed to the dispersive delay, because the time delay at a given frequency $\nu$ follows a $\nu^{-2}$ law and the column density of free electrons inferred from the time delay is roughly consistent with the theoretical predictions [accumulated contributions from MK, IGM and host galaxy  \citep{chatterjee17}]. The time-frequency structure of the FRB 121102 sub-pulses, however, emerges after correcting for dispersive delays. Therefore, the time delay between sub-pulses are largely reduced to the order of milliseconds or even sub-milliseconds, which is very advantageous for further improving the accuracy of a basic physical analysis. Moreover, it has been proposed that the observed downward drifting of the sub-pulse frequency is more likely intrinsic, namely a generic geometrical effect within the framework of coherent curvature radiation by bunches of electron- positron pairs in the magnetosphere of a neutron star \citep{wang19}. If this is the case, the constraints on the WEP and the photon mass would become even tighter. 

\section{acknowledgments}

This work is supported by the National Natural Science Foundation of China (NSFC) under Grant No. 11690024, 11722324, 11603003, 11633001, 11725314, 11603076, U1831122 the Strategic Priority Research Program of the Chinese Academy of Sciences, Grant No. XDB23040100 and the Fundamental Research Funds for the Central Universities. WYW acknowledges the support from MoST grant 2016YFE0100300, NSFC under Grant No. 11633004, 11473044, 11653003, and the CAS grants QYZDJ-SSW-SLH017. PM acknowledges the Eberly Foundation.

\end{document}